\documentclass[%
reprint,superscriptaddress,
amsmath,amssymb,
aps,
prl,
showpacs,
]{revtex4-1}
\usepackage{color}
\usepackage[usenames,dvipsnames,svgnames,table]{xcolor}
\usepackage{amssymb}
\usepackage{graphicx}
\usepackage{dcolumn}
\usepackage{bm}
\usepackage{hyperref}
\hypersetup{backref,colorlinks=true,linkcolor=NavyBlue,citecolor=BrickRed}

\begin{document}

\title{Hund-enhanced electronic compressibility in FeSe and its correlation with T$_c$}

\author{Pablo Villar Arribi}
\affiliation{European Synchrotron Radiation Facility, 71 Avenue des Martyrs, F-38000 Grenoble, France}

\author{Luca de' Medici}
\affiliation{Laboratoire de Physique et Etude des Mat\'eriaux, UMR8213 CNRS/ESPCI/UPMC, Paris, France} 
\affiliation{European Synchrotron Radiation Facility, 71 Avenue des Martyrs, F-38000 Grenoble, France} 
\email{demedici@esrf.fr}

\date{\today}

\begin{abstract}
We compute the compressibility of the conduction electrons in both bulk orthorhombic FeSe and monolayer FeSe on SrTiO$_3$ substrate, including dynamical electronic correlations within
slave-spin mean-field + density-functional theory. Results show a zone of enhancement of the electronic compressibility crossing the interaction-doping phase diagram of these compounds
in accord with previous simulations on iron pnictides and in general with the phenomenology of Hund's metals. Interestingly at ambient pressure FeSe is found slightly away from the zone
with enhanced compressibility but moved right into it with hydrostatic pressure, while in monolayer FeSe the stronger enhancement region is realized on the electron-doped side.
These findings correlate positively with the enhancement of superconductivity seen in experiments, and support the possibility that Hund's induced many-body correlations boost
superconductive pairing when the system is at the frontier of the normal- to Hund's-metal crossover.
\end{abstract}
\maketitle

Iron-based high-T$_c$ superconductors (IBSC) were discovered ten years ago~\cite{kamihara2008iron}, but much has yet to be clarified about their physics. Many of these materials exhibit
long-range magnetic order which is suppressed in favor of a superconducting phase when the compound is doped or put under pressure~\cite{Paglione_review, IBSC_springer, martinelli_phase_diagrams},
suggesting that spin-fluctuation mediated interactions could play an important role on the pairing mechanism for superconductivity~\cite{fernandes_nature_spin_fluc}.

However notable exceptions to this phenomenology pose serious questions about its general validity. FeSe is a striking one. Indeed this compound at ambient pressure is a non-magnetic metal
undergoing a nematic instability~\cite{Boehmer_PRL_Tetra_Ortho} (with concomitant tetragonal-to-orthorhombic structural transition) below $\sim$90 K. Superconductivity arises below $\sim$9 K,
apparently unrelated to magnetism, and is enhanced when applying hydrostatic pressure. The maximum T$_c\sim$37 K is reached around 7-9 GPa in this
orthorhombic phase~\cite{Margadonna, medvedev_fese_pressure}, beyond which the decrease of T$_c$ seen in experiments is accompanied and possibly favored by the coexistence of different
crystallographic phases (tetragonal, hexagonal, orthorhombic,...)~\cite{Svitlyk_FeSe_pressure_struct}.

Even more surprisingly a single layer of FeSe deposited on a SrTiO$_3$ substrate (FeSe/STO) shows the highest T$_c$ ($>$65 K, and perhaps even over 100 K)~\cite{FeSe_STO_100} reported thus far in IBSC.
This is further remarkable since the spin fluctuations potentially responsible for the high-Tc superconductivity arise quite naturally out of nesting between roughly equally-sized hole and electron
Fermi pockets, typically found in the band structures of IBSC which are compensated semimetals. However FeSe/STO appears to be electron doped\cite{Tan_DLFeng-el_doped-STO},
since ARPES measures only show electron pockets, thus questioning the spin-fluctuation scenario. 

A missed ingredient that might contribute substantially to superconductive pairing in this case or in general is phonons even if they were very early ruled out as the main source of pairing
in iron pnictides~\cite{Boeri_FeSC_noPhonons,Haule_LaOFFeAs_PRL2008}. They have been called into the game again for FeSe/STO: in particular phonons of
the substrate could be more effective in enhancing the superconductivity of the monolayer than those of FeSe itself~\cite{DHLee_FeSeMono_eph_STO,Rademaker-FeSe_STO_eph}, but it might also be
that this electron-phonon coupling is substantially screened by the same electrons of FeSe, as recently claimed in Ref.~\onlinecite{Zhou_Millis-STO_Phonons_FeSe_Mono}.

In this work we study theoretically the compressibility of conduction electrons in FeSe (both bulk and monolayer), and we highlight an enhancement of this quantity that correlates positively
with the enhancement of the T$_c$ found experimentally. 

This enhanced compressibility is the outcome of many-body correlations due to the interaction among the electrons. 
Indeed correlations need to be included in the ab-initio simulations of IBSC~\cite{Shorikov2009, aichhorn_silke_fese, yin_haule_kotliar_ibsc_nature, werner2012satellites,
misawa_imada_mott_ibsc, qimiao_si_htc_ibsc_review}  to fit most experimental results and FeSe is thought to be one of the most correlated materials in this
family~\cite{aichhorn_silke_fese, yin_haule_kotliar_ibsc_nature,Watson-HubbardBands_FeSe,Evtushinsky_Borisenko-HubBandsFeSe}. In particular Hund's coupling, i.e. intra-atomic exchange,
plays a fundamental role, so much that IBSC are considered a paradigm for ``Hund's metals''~\cite{yin_haule_kotliar_ibsc_nature,Antoine_Luca_Jernej_Hund}. Among the main defining features
of this kind of phase are: dominant high-spin configurations causing a large fluctuating local moment in the paramagnetic metal~\cite{lafuerza_ibsc_moments}, and large and strongly
orbitally-differentiated correlation strengths~\cite{Luca_PRL_IBSC_2014, yin_haule_kotliar_ibsc_nature}. Recently~\cite{demedici_el_comp} it was shown that an enhanced electronic compressibility
(culminating in a divergence) ubiquitously accompanies the cross-over between the Hund's metal and the normal metal in multi-orbital Hubbard models in presence of Hund's coupling.
This crossover departs from the Mott transition that is found at half-filling at rather low interaction strength~\cite{Ishida_Mott_d5_nFL_Fe-SC,Luca_PRL_IBSC_2014,Fanfarillo_Hund,misawa_imada_mott_ibsc}
and extends to finite doping and larger interaction strengths. In a realistic simulation~\cite{demedici_el_comp} of the "122" family of IBSC (BaFe$_2$As$_2$ and similar compounds)
the tip of this region of enhanced compressibility was shown to extend into the region where high-Tc superconductivity and the other instabilities happen experimentally,
and it was advanced that the enhanced quasiparticle interactions causing the enhanced electronic compressibility might also be the cause of enhancement of all the other instabilities,
including superconductivity (in line with \onlinecite{Misawa_LaFeAsO} and in the general framework of Refs. \cite{EmeryKivelson,GrilliRaimondi_IntJModB,CDG_PRL95}).
Here we show that the enhancement of the electronic compressibility is also found in a realistic simulation of FeSe under pressure and of electron-doped FeSe monolayer on STO,
which are the cases of maximum T$_c$ in chalcogenides, thus corroborating this suggestion. 	

The common block of all IBSC are the layers of buckled planes made of Fe atoms with the ligands (pnictogens or chalcogens) located above and below alternatively.
It is in these planes where superconductivity is thought to occur. Indeed in all IBSC 5 bands of mainly Fe-$3d$ character cross the Fermi level, with a total bandwidth around ~4 eV.
FeSe is the simplest of all IBSC as it is composed only of a stacking of such planes.

We thus model the conduction electrons here with a 5-orbital Hubbard-Kanamori Hamiltonian, $\hat{\mathcal{H}}-\mu\hat{N}=\hat{\mathcal{H}}_0+\hat{\mathcal{H}}_{int}-\mu\hat{N}$, where $\mu$ is
the chemical potential. This Hamiltonian includes a non-interacting part
\begin{equation}
\hat{\mathcal{H}}_0=\sum_{i\neq j,m,m',\sigma}t^{mm'}_{ij}d^{\dagger}_{im\sigma}d_{jm'\sigma}+\sum_{i,m,\sigma}\varepsilon_m\hat{n}_{im\sigma},
\label{eq:H0}
\end{equation}
where $d^{\dagger}_{im\sigma}$ creates an electron with spin $\sigma$ in orbital $m=1,...,5$ on the site $i$ of the lattice, and
$\hat{n}_{im\sigma}=d^{\dagger}_{im\sigma}d_{im\sigma}$ is the number operator.
The hopping integrals $t^{mm'}_{ij}$ and the on-site orbital energies $\varepsilon_m$ are obtained by means of a
tight-binding parametrization of the bare band structure, which is calculated within the Density Functional
Theory (DFT) framework with the code {\sc Wien2k}~\cite{WIEN2k} using the GGA-PBE exchange-correlation functional~\cite{PBE}.
This parametrization is written in a basis of maximally-localized Wannier functions~\cite{Wannier_review}
including only conduction bands of mainly Fe-3$d$ character, and is computed using the {\sc Wannier90} code~\cite{Wannier90}.
For these calculations, the lattice parameters and atomic positions for the bulk in the orthorhombic phase (which is the one realized at low temperature in the range
of pressures of interest here~\cite{Margadonna, medvedev_fese_pressure,kumar_fese}) are taken from Ref.~[\onlinecite{Margadonna}], and for the monolayer we fix
the $a$ and $b$ lattice parameters to those of STO ($a=b=3.905$ \AA{})
and $z_{\mathrm{Se}}$ (the height of the ligand) is taken from Ref.~[\onlinecite{mandal_haule_fese_correl}]. 
The many-body interacting part of the Hamiltonian reads:
\begin{equation}
\label{eq:Hint}
\begin{split}
\hat{\mathcal{H}}_{int}&=U\sum_{m}\hat{n}_{m\uparrow}\hat{n}_{m\downarrow}+U'\sum_{m\neq m'}\hat{n}_{m\uparrow}\hat{n}_{m'\downarrow}\\
&+(U'-J)\sum_{m<m',\sigma}\hat{n}_{m\sigma}\hat{n}_{m'\sigma}\\
&-J\sum_{m\neq m'}d^{\dagger}_{m\uparrow}d_{m\downarrow}d^{\dagger}_{m'\downarrow}d_{m'\uparrow}\\
&+J\sum_{m\neq m'}d^{\dagger}_{m\uparrow}d^{\dagger}_{m\downarrow}d_{m'\uparrow}d_{m'\downarrow}
\end{split}
\end{equation}
where $U$ is the local-on-site Coulomb repulsion, and $J$ the Hund's coupling.  
Customarily we choose $U'=U-2J$ and we drop the last two terms in eq.(\ref{eq:Hint}) (spin-flip and pair-hopping respectively) which moreover
need extra approximations to be treated exactly in our method of choice to deal with many-body correlations, Slave-Spins Mean-Field Theory (SSMFT)~\cite{luca_massimo_ssmft}.
This is a very convenient approach since it describes by construction a Fermi liquid, which is the behavior shown by IBSC at low temperature~\cite{RullierAlbenque_Fermi_liquid}, 
it successfully describes the orbital differentiation of these materials~\cite{Luca_PRL_IBSC_2014}, and accurately predicts the Sommerfeld coefficient of the 122 family.~\cite{hardy_luca_122}
We choose $U=4.2$ eV for FeSe (although several scans in $U$ are performed) and we fix $J/U=0.2$. These values are obtained by ab-initio constrained random-phase
approximation calculations~\cite{Miyake} (cRPA).

It is worth signalling that albeit FeSe is a much studied material, some basic details about it must still be understood. 
For instance experiments disagree among them on the precise value of $z_{\mathrm{Se}}$~\cite{Margadonna,kumar_fese}, particularly for FeSe under pressure, and theoretical
simulations also give somewhat different values~\cite{mandal_haule_fese_pressure}. Details of the DFT band structure indeed turn out to be quite sensitive to this parameter. 
Also it is now accepted that standard DFT does not provide a quantitatively accurate Fermi surface for IBSC. The basic compensated semi-metal character with both hole- and electron- pockets is
indeed correctly predicted but the size of the pockets is in all cases (and for FeSe severely) too large compared to experiments. However here we will concentrate on the many-body physics,
that is dominated by the local energetics determined by $\hat{\mathcal{H}}_{int}$. It is indeed also strongly influenced by the bare electronic structure but this happens
mainly through local (i.e. $k$-averaged) quantities, like the crystal-field splitting of orbital energies, the (total and orbital-resolved) kinetic energy, etc.
We thus expect it to be much more robust to changes and eventual improvements in $\hat{\mathcal{H}}_0$ than the sheer fermiology. 

SSMFT describes the Fermi-liquid low-temperature paramagnetic metallic phase of this model as a quasiparticle Hamiltonian~\cite{luca_massimo_ssmft}
\begin{equation}
\hat{\mathcal{H}}_{QP}=\!\!\!\!\!\!\!\!\sum_{i\neq j,m,m',\sigma}\!\!\!\!\!\!\!\!\sqrt{Z_mZ_{m'}}t^{mm'}_{ij}f^{\dagger}_{im\sigma}f_{jm'\sigma}+
\sum_{i,m,\sigma}(\varepsilon_m-\lambda_m)\hat{n}^f_{im\sigma},
\label{eq:Heff}
\end{equation}
where   $f^{\dagger}_{im\sigma}=\sum_k e^{ik\cdot r_i} f^{\dagger}_{km\sigma}/\sqrt{{\cal N}_{sites}}$ and $f^{\dagger}_{km\sigma}$
creates a quasiparticle with corresponding quantum numbers. The number of quasiparticle equals the number of particles owing to the Luttinger theorem and thus
\begin{equation}
n_f\equiv \sum_{km\sigma}\langle f^\dagger_{km\sigma}f_{km\sigma} \rangle=\int^\mu d\varepsilon D^*(\varepsilon)=n,
\label{eq:density}
\end{equation}
where $n$ is the average electron density. $D^*(\varepsilon)$ is the renormalized (quasiparticle) density of states (DOS), and the renormalization 
due to the interaction eq. (\ref{eq:Hint}) in SSMFT is brought in by the factors $Z_m$ (that act as inverse mass enhancements factors) and $\lambda_m$ (that shift the on-site energy).
These renormalization factors are calculated in a set of self-consistent mean-field equations that involve the auxiliary slave-spin variables~\cite{luca_massimo_ssmft}.
They thus depend on all the physical parameters of the problem in a non-trivial way. Importantly this means that the quasiparticle model is not (for a given set of interaction parameters $U,J$)
just a "rigid" renormalized band structure by respect to i.e. filling or temperature, but a structure that changes when these parameters change. 
We are for instance here interested in the electronic compressibility $\kappa_{el}=\frac{d n}{d \mu}$. In this mean field it reads, deriving eq. (\ref{eq:density}):
\begin{equation}
\kappa_{el}=\frac{D^\ast(\mu)}{1-\int^\mu d\varepsilon \frac{dD^\ast}{dn}(\varepsilon)}
\label{eq:comp_Fermi}
\end{equation}
where it is clear that the renormalized "rigid" band structure value $D^\ast(\mu)$ is corrected by the expression at the denominator due to the change
in the band structure with the filling, that thus plays the role of the Landau parameter $F^s_0$ of an isotropic Fermi liquid~\cite{nozieres}.

An enhanced compressibility can thus stem from a strong renormalization of the DOS (i.e. a large $D^\ast(\mu)$) or from a Landau
parameter $-\int^\mu d\varepsilon \frac{dD^\ast}{dn}(\varepsilon)$ (usually small and positive) becoming negative and approaching -1, or from both.

\begin{figure}[ht]
    \begin{center}
       \includegraphics[height=5cm]{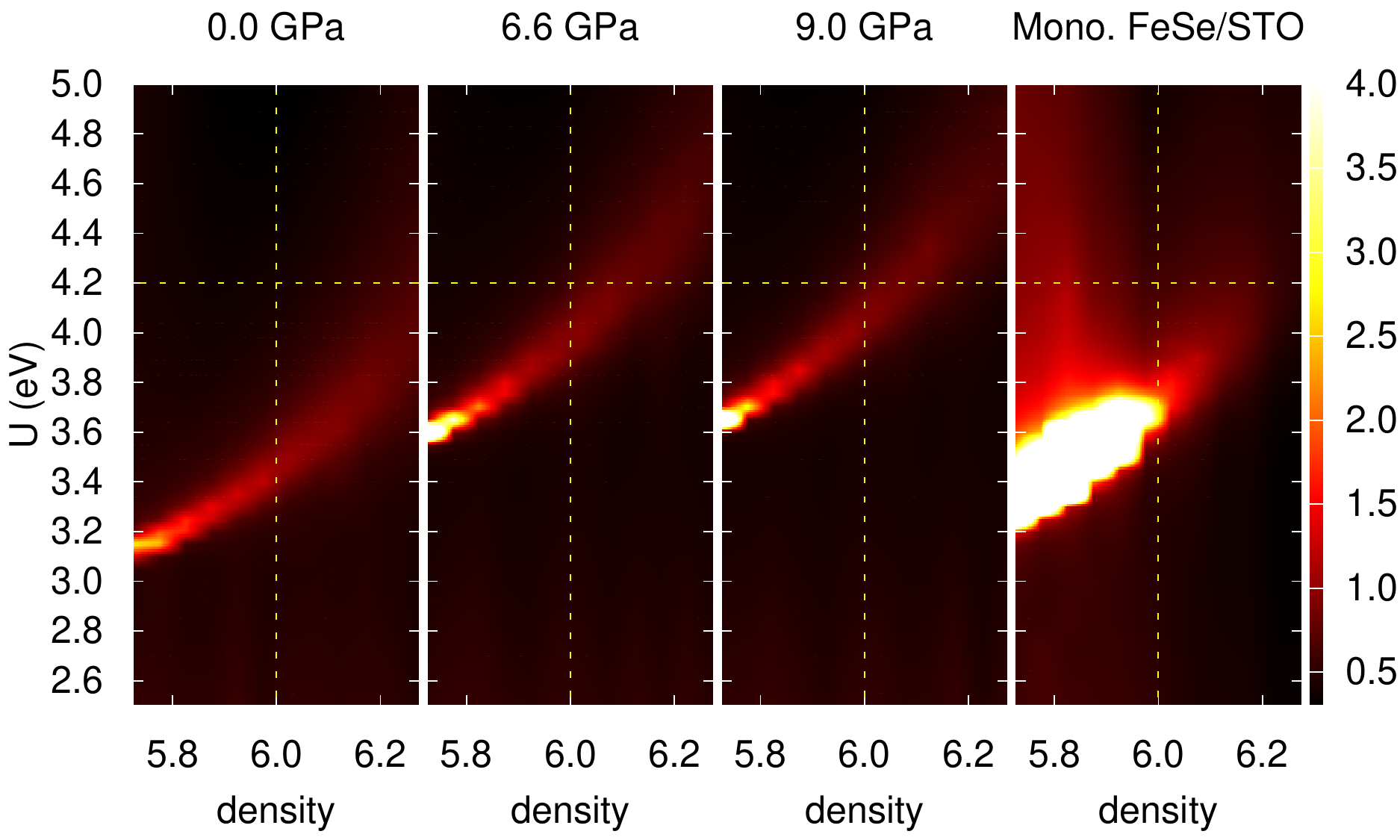}
       \caption{Color map of the electronic compressibility $\kappa_{el}=dn/d\mu$ (in color scale) of FeSe.
       The brighter the color, the larger the electronic compressibility, until reaching the white color, which denotes a divergence in the electronic compressibility.
       The stair structure that can be observed is unphysical and corresponds to the numerical discretization of the derivative. 
       From left to right: electronic compressibility for 0.0, 6.6 and 9.0 GPa cases and for a monolayer of FeSe over a substrate of STO. 
       The vertical yellow dashed lines represent the stoichiometric filling ($n=6.0$) and the horizontal ones our estimated value
       of $U=4.2$ eV for this system. The crossing point locates the stoichiometric compound in this U-filling plane.}
       \label{fig:map}
    \end{center}
 \end{figure}

Our main result is shown in Fig. \ref{fig:map}.
We have calculated the electronic compressibility $\kappa_{el}$ of FeSe for a range of
dopings in the vicinity of the stoichiometric compound ($n=6.0$), and for different values
of the Coulomb repulsion $U$. Each of these sets of calculations was performed for FeSe
at three different values of the hydrostatic pressure of 0.0, 6.0 and 9.0 GPa, and also
for a monolayer of FeSe on top of a substrate of STO.

\begin{figure}[ht]
    \begin{center}
       \includegraphics[width=8.2cm]{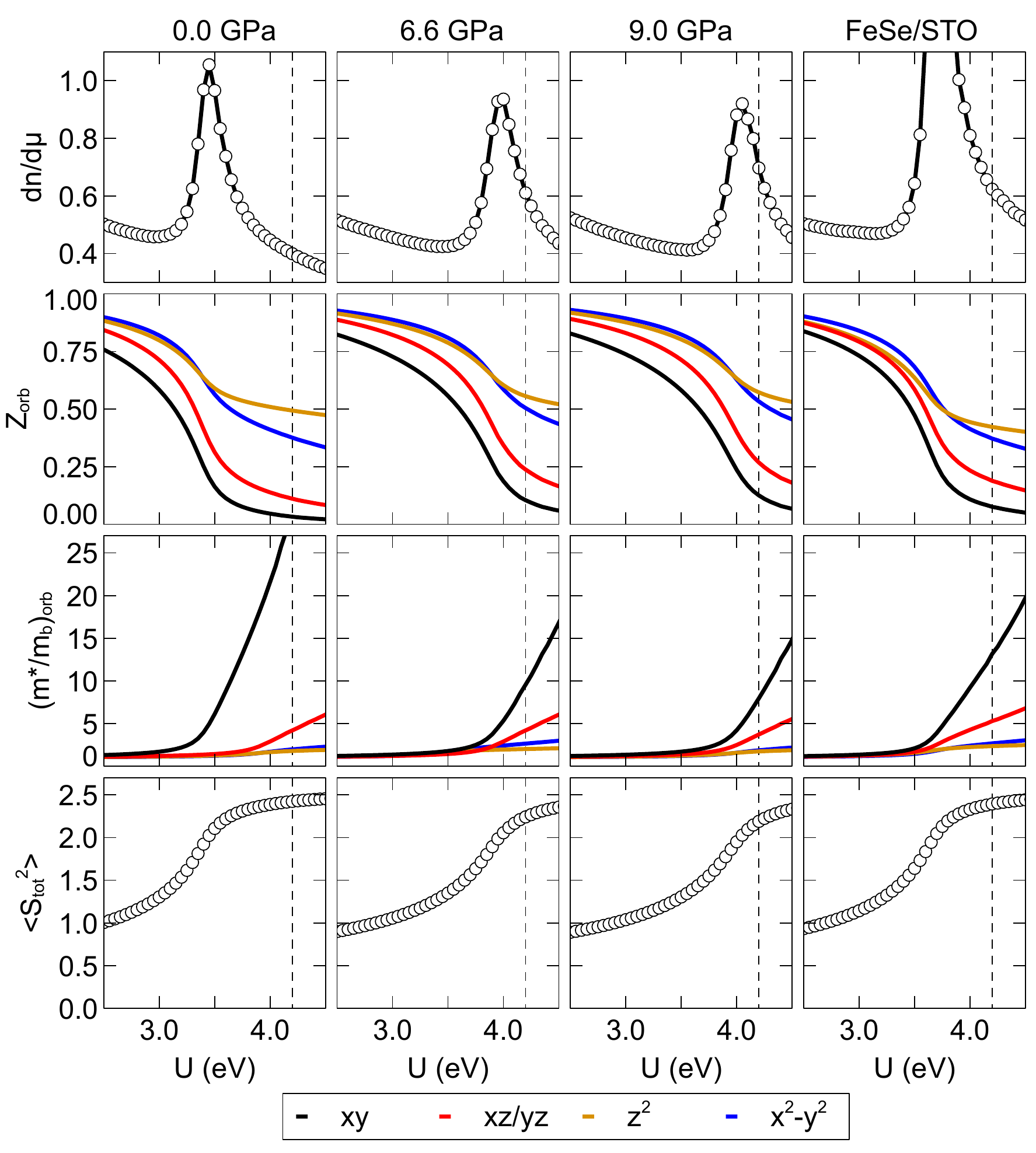}
       \caption{Results for bulk FeSe for three different values of applied hydrostatic pressure and for monolayer FeSe/STO.
       Upper panels: electronic compressibility as a function of the on-site Coulomb interaction $U$.
       Middle-upper panels: quasiparticle weights of the different orbitals as a function of $U$.
       Middle-lower panels: mass enhancement of the different orbitals ($1/Z_m$) as a function of $U$.
       Lower panels: total local spin-spin correlation function as a function of $U$.
       All calculations are performed for or $n=6.0$ filling and $J/U$=0.2}
       \label{fig:comp_mass_U}
    \end{center}
 \end{figure}

Our calculations clearly show that the enhancement in the electronic compressibility with a "moustache" shape is present in the system (brighter
region in the color maps) analogously to the case of BaFe$_2$As$_2$~\cite{demedici_el_comp}.
But unlike the latter compound, which happens to be on top the region of enhancement, the realistic parameter values for stoichiometric bulk FeSe (dotted lines)
are located at some distance from the region of enhanced compressibility. Remarkably however, increasing pressure moves the enhancement region to the physical values. 
This is further illustrated by the upper panel of Fig. \ref{fig:comp_mass_U}. Thus assuming realistically that the interaction strength is not sensibly modified by the applied pressure,
one sees that for 6-9 GPa the enhancement region has basically reached (although not completely) the physical parameters.
This is remarkable in that the same trend is observed in the experimental T$_c$, which tops in the same range of pressures,
before crystallographic changes intervene~\cite{medvedev_fese_pressure,Svitlyk_FeSe_pressure_struct}.

It is tempting to interpret our results very literally: the considerable increase of a factor of 2 or larger that we find in the electronic compressibility between 0 and $\sim$ 9 GPa 
might result in a considerable enhancement of the superconductive pairing, whereas a further enhancement of $\kappa_{el}$ makes this phase too susceptible to transitions towards other
crystal structures and cuts off a further enhancement of T$_c$. More realistically the outlined caveats on the knowledge of the exact atomic positions under pressure,
error bars on the estimates of the interaction parameters and the known (albeit somewhat under control) inaccuracy of the details of the DFT band structure suggest to consider
only the main trends here. In this respect our result is very robust: there is a region of compressibility enhancement (culminating in a divergence below some filling) in
the phase diagram of FeSe that is entered when pressure in the correct experimental range is considered in our simulations.

The above scenario is determined by the fact that, as mentioned in the introduction, the moustache-shaped zone of compressibility enhancement is found in both
models and realistic simulations of Hund's metals, in their low-temperature Fermi-liquid phase. It departs from the Mott transition point at half filling and
extends at finite dopings. This region parallels the universal cross-over~\cite{demedici_Hunds_metals} between normal (at low-U and large doping) and Hund's metal (at large U and small doping).

FeSe is believed to be more correlated than BaFe$_2$As$_2$~\cite{yin_haule_kotliar_ibsc_nature} and thus it is plausible that if the latter lies on top of the
crossover (as found in Ref.~\onlinecite{demedici_el_comp}) , the former might be well inside the Hund's metal region.
In our results (plotted in Fig. \ref{fig:comp_mass_U}) this fact is clearly indicated by FeSe at ambient pressure showing the hallmarks
of the Hund's metals {\it i)} large fluctuating total local magnetic moment $\langle S_z \rangle$ (lower panels), {\it ii)} orbitally differentiated mass
enhancements (middle-lower panels), {\it iii)} strong correlations and low Fermi-liquid coherence scales (due to the low quasiparticle weights - middle-upper panels - corresponding to the
large values of the mass enhancement for
the conduction electrons, of main orbital character $xy$, $xz$,$yz$). The crossover in these quantities towards their typical behavior in a more conventional
metal is clearly visible at lower U than the "realistic" value 4.2 eV. As expected the compressibility enhancement (upper panels) is shown to track this crossover.

With applied hydrostatic pressure (which indeed increases the bandwidth while leaving the interaction basically untouched, so its effect can be inferred by
the results at reduced interaction strength) the crossover moves towards higher U and approaches the realistic value. In our calculations basically at the
pressure 6-9 GPa the compound is predicted (within all the previously outlined caveats) almost on top of the Hund's-to-normal crossover. Indeed we see
from Fig. \ref{fig:comp_mass_U} that at the crossover $\langle S_z \rangle$ is expected to reduce rapidly and that the enhancement of the masses should
go back to moderate and with small differentiation among the different orbitals. Consequently the Fermi-liquid coherence scale is expected to grow much larger.

 Some experimental support for this scenario can be found in the literature. 
{\it i)} The estimate of the local paramagnetic moment by X-ray Emission Spectroscopy (XES) is seen to drop monotonously in the range of pressures
0-9 GPa~\cite{Kumar_XES-FeSe}(before the system undergoes a change of structure producing an even higher value for the moment~\cite{Lebert_Toulemonde-XES_FeSe}.
{\it ii)} Orbital selective correlations have been directly reported from ARPES or Quantum Oscillations in all Fe-chalcogenides~\cite{Yi_Universal_OSM_Chalcogenides} and
in FeSe in particular~\cite{Watson_Coldea_ARPES}, in the normal phase. Remarkably, it was shown lately by quasiparticle interference on STM measures that the
superconducting gap shape cannot be explained without including heavily orbitally-differentiated quasiparticle weights~\cite{Sprau_Davis-OrbSel_Pairing_FeSe,Kostin_Davis-OrbSel_normal_FeSe}.
It is quite safe to conclude that FeSe lies in a regime of strong orbital differentiation of the correlation strength as predicted
theoretically~\cite{yin_haule_kotliar_ibsc_nature,Lanata_Luca_FeSe_GA} and in agreement with the general mechanism outlined in Ref.~\onlinecite{demedici_OSMT_deg_lift_prl}.
{\it iii)} A remarkable crossover is found in the resistivity around 350K~\cite{Karlsson_Toulemonde-FeSe_crossover350K}. While at low temperature the behavior is metallic,
after a shoulder located around 350K it starts decreasing with temperature, signalling a cross-over towards bad-metallic/semiconducting behavior.
This fact is readily interpreted as a low coherence scale of the metallic carriers.

We can thus conclude that the calculations at the estimated values for the interactions seem to reproduce correctly the Hund's metal behavior of FeSe found in experiments,
and the prediction of the zone of enhanced compressibility at 6-9 GPa can be deemed robust.

The parallelism between the theoretical enhancement of the electronic compressibility we found here and the experimental enhancement of the
superconducting T$_c$ can support the view of a mechanism for enhancing superconductivity based on local electronic interactions (and in particular Hund's coupling),
as proposed in Ref.~\onlinecite{demedici_el_comp}. Indeed, following formula (\ref{eq:comp_Fermi}), $\kappa_{el}$ can be large (or even diverge) because of a large
numerator or a small denominator. The first case simply indicates strong quasiparticle renormalization. The second, which is the case realized in this kind
of instabilities~\cite{demedici_el_comp} indicates attractive forces in the particle-hole channel that can lead to a negative scattering amplitude in the particle-particle
channel, and thus to superconductivity~\cite{GrilliRaimondi_IntJModB}.

Furthermore some electron-boson vertices can also be enhanced by electron-electron interactions, like for instance the density vertex (relevant for Holstein electron-phonon coupling). Indeed the 
following Ward identity holds~\cite{Grilli_El-Ph} for the renormalized density vertex $\Lambda(q\rightarrow0,\omega=0)=\frac{1}{Z(1+F^s_0)}$ in an isotropic Fermi liquid.
One can see how the vertex is renormalized in the same way as the electronic compressibility, thus leading to
a enhanced effective interaction strength which may trigger any particular mechanism mediated by this kind of interaction.

Let's now turn to the case of monolayer FeSe. In the case of FeSe/STO, as visible in the rightmost panel in Fig. \ref{fig:map}, in our calculations the enhancement
region is much larger and the enhancement itself is more intense overall.  This can be correlated positively with the enhanced experimental T$_c$ of the monolayer, in the same spirit as above.
A peculiar shape is also noticeable, of the enhancement region, that seems to "bifurcate" for U $\gtrsim$ 3.8eV in a branch that extends to electron doping (density values n$\simeq 6.1\div 6.2$)
and another to hole doping (n$\simeq 5.7\div 5.9$). By analyzing the renormalized DOS at the Fermi level $D^*(\mu)$ one can show that the hole-doping branch is due to
an enhanced structure in $D^*(\mu)$, while the one at electron doping is not, and there $\kappa_{el}$ is thus enhanced by the denominator in formula (\ref{eq:comp_Fermi}).
This means that the enhancement branch at electron doping is the genuine continuation of the "moustache" structure, carrying over all the physical considerations done so
far about it (indeed the behavior of all the quantities analyzed in Fig. \ref{fig:comp_mass_U} for FeSe/STO parallels the corresponding ones in FeSe).
This again correlates positively with experiments in that FeSe/STO with the enhanced T$_c$ is electron-doped~\cite{Tan_DLFeng-el_doped-STO}.
It might also be worth to stress here that the STO substrate has a very high dielectric constant which might contribute to the screening of the electronic interactions in FeSe,
so that the actual value of $U$ for the Fe-$3d$ electrons in this system could eventually become a bit lower. 

It is worth also signalling that the enhancement of the compressibility reported here points to a non-negligible role played by local vertex corrections,
as explicitly shown by the mentioned Ward identity. If the present scenario is realized then, one might reconsider the suppression of electron-phonon coupling due to the screening
of FeSe conduction electrons estimated in Ref.~\onlinecite{Zhou_Millis-STO_Phonons_FeSe_Mono}, which was done precisely neglecting vertex corrections.
Electron-phonon coupling might actually be boosted in the rather narrow region corresponding to the enhanced compressibility, as also
calculated in Ref.~\onlinecite{mandal_haule_fese_pressure} for bulk FeSe, and thus contribute substantially to the high-temperature superconductivity.

In summary, we have solved a multi-orbital Hubbard Hamiltonian for FeSe bulk at different pressures
and for a monolayer of FeSe within the SSMFT framework, and studied the electronic compressibility 
$\kappa_{el}=dn/d\mu$ of each of these systems. At ambient pressure, an enhancement of $\kappa_{el}$
is found in the doping-interaction plane but is slightly off from the realistic value of the on-site Coulomb repulsion $U$ for this compound at
the stoichiometric filling of $n=6.0$. This enhancement region (which at lower electron densities - i.e. strong hole doping - turns
into a divergence, signalling an instability region towards phase separation there) is moved even closer to the realistic parameters for FeSe when
pressure is increased, showing an analogous enhancement as the critical temperature ($T_c$) of FeSe. The largest electronic compressibility is finally achieved in the range
of pressures in which FeSe presents a higher $T_c$ (around 9 GPa). These trends are consistent in the case of a monolayer of FeSe, where the instability region is larger
and the enhancement of $\kappa_{el}$ more intense overall, and extends to electron doping, reproducing the trend of the experimental $T_c$.

\acknowledgments	

The authors are supported by the European Commission through the ERC-StG2016, StrongCoPhy4Energy, GA No724177.

%
\end{document}